\documentclass[%
 aip,
 amsmath,amssymb,
 reprint,%
]{revtex4-1}

\usepackage{graphicx}% Include figure files
\usepackage{dcolumn}% Align table columns on decimal point
\usepackage{bm}% bold math

\usepackage[utf8]{inputenc}
\usepackage[T1]{fontenc}
\usepackage{mathptmx}
\usepackage{etoolbox}
\usepackage{color,soul}
\usepackage{siunitx} % enables \SI{}{}
\usepackage{setspace}
\usepackage{textcomp}

\makeatother

\begin{document}

\preprint{AIP/123-QED}

\title[Polarization Study of Single Color Centers in Aluminum Nitride]{Polarization Study of Single Color Centers in Aluminum Nitride}

\author{J.K. Cannon}
 \affiliation{School of Engineering, Cardiff University, Queen's Buildings, The Parade, Cardiff, UK, CF24 3AA}
   \affiliation{Translational Research Hub, Maindy Road, Cardiff, CF24 4HQ, UK}
\author{S.G. Bishop}
 \affiliation{School of Engineering, Cardiff University, Queen's Buildings, The Parade, Cardiff, UK, CF24 3AA}
   \affiliation{Translational Research Hub, Maindy Road, Cardiff, CF24 4HQ, UK}
\author{J.P. Hadden}
\affiliation{School of Engineering, Cardiff University, Queen's Buildings, The Parade, Cardiff, UK, CF24 3AA}
  \affiliation{Translational Research Hub, Maindy Road, Cardiff, CF24 4HQ, UK}
\author{H.B. Ya\u{g}c{\i}}
\affiliation{School of Engineering, Cardiff University, Queen's Buildings, The Parade, Cardiff, UK, CF24 3AA}
  \affiliation{Translational Research Hub, Maindy Road, Cardiff, CF24 4HQ, UK}
\author{A.J. Bennett} 
 \affiliation{School of Engineering, Cardiff University, Queen's Buildings, The Parade, Cardiff, UK, CF24 3AA}
  \affiliation{Translational Research Hub, Maindy Road, Cardiff, CF24 4HQ, UK}
   \affiliation{School of Physics and Astronomy, Cardiff University, Queen's Buildings, The Parade, Cardiff, CF24 3AA, UK}
 \address{Author to whom correspondence should be addressed: BennettA19@cardiff.ac.uk}

\begin{abstract}
 Color centers in wide-bandgap semiconductors are a promising class of solid state quantum light source,  many of which operate at room temperature. We examine a family of color centers in aluminum nitride which emit close to \SI{620}{\nano \meter}. We present a technique to rapidly map an ensemble of these single photon emitters, identifying all emitters, not just those with absorption dipole parallel to the laser polarization. We demonstrate a fast technique to determine their absorption polarization orientation in the c-plane, finding they are uniformly distributed in orientation, in contrast to many other emitters in crystalline materials.
\end{abstract}

\flushbottom
\maketitle

Color centers (CCs) are emissive point structures in a crystal consisting of combinations of impurities, vacancies and lattice defects which at high-density can give a crystal color. When a single center is isolated, the photon statistics often demonstrate anti-bunching. CCs with internal energy levels embedded deep within the bandgap of their semiconductor host can emit at room-temperature. The most investigated example is the negatively charged nitrogen vacancy (NV$^-$) complex in diamond\cite{Kurtsiefer2000StablePhotons} which shows great promise as a room temperature light source for quantum communications \cite{Ju2020PreparationsDiamond}, bio-compatible quantum-sensor \cite{Maze2008}, long-lived spin memory \cite{Balasubramanian2009SpinMemory} and as a platform for tests of fundamental quantum physics \cite{Hensen2015}. In recent years, CCs have been found in a large range of other wide-band-gap materials including silicon carbide \cite{Klimov2014} and hexagonal boron nitride \cite{Martinez2016EfficientCrystal, Camphausen2020ObservationTemperature}.

Isolated CCs in the group III-nitrides have been less well investigated, with a small number of works reporting their presence in gallium nitride (GaN)\cite{Berhane2017, Zhou2018RoomRange, Bishop2022} and aluminum nitride (AlN)\cite{Bishop2020Room-TemperatureNitride, Xue2020Single-PhotonFilms, Xue2020} emitting in the visible and near infrared. The prevalence of these semiconductors in high power electronics and solid state lighting makes this an especially promising platform to investigate, with an established route to cost-effective and commercial-scale fabrication, and epitaxial material available at low cost. AlN is a wurzite semiconductor with a direct band-gap of \SI{6.2}{\electronvolt} and a high refractive index of 2.14 at \SI{600}{\nano\meter} \cite{GaetaPhotonic-chip-basedCombs}. Its wide transparency window, second and third order optical non-linearity \cite{Li2021AluminiumReview} have motivated the development of waveguide-coupled devices  \cite{Chen2021DevelopmentSpectrum, Liu2015OpticalOptics, Xiong2012} including cavity structures \cite{Sam-Giao2012HighWaveguide}. This processing technology could be adapted to integrate CCs in photonic integrated circuits for applications in quantum technologies.

In previous publications, density functional theory has been used to predict emission from several CCs in AlN, namely the anti-site nitrogen vacancy complex ({N$_{\text{Al}}$V$_{\text{N}}$}) at \SI{712}{\nano \meter}\cite{Xue2021}, the divacancy ({V$_{\text{Al}}$V$_{\text{N}}$}) at \SI{867}{\nano \meter}\cite{Xue2021} and negative vacancy (V$_{\text{N}}^{-}$) at \SI{443} - \SI{517}{\nano \meter} \cite{Varley2016}. The conditions to form these CCs are not well understood, but one report has shown CCs in GaN are correlated with the density of threading dislocations but are not correlated spatially \cite{Nguyen2019}, which may point to a common cause. Other works have pointed to the polarity of the crystal playing a role in the epitaxy of material containing CCs in GaN \cite{Nguyen2021}. However, the small number of experimental reports on isolated CCs in AlN means there is little evidence CCs in GaN and AlN are related, even if they share some optical characteristics and exist in similar semiconductors.

Here we report our studies on isolated CCs in a commercially-sourced AlN-on-sapphire sample that has a convenient density of CCs with zero-phonon lines clustered around to \SI{620}{\nano \meter}. To learn more about the AlN CC's physical origin we investigate the ensemble distribution of absorption dipoles. We find that, despite the crystalline nature of the film, there is no preferred absorption dipole orientation direction for the CCs. Detailed study of the temporal, spectral and photo-physical properties of a small number of CCs provides an insight into their internal energy levels and physical structure.

\begin{figure}
    \centering
    \includegraphics[scale = 1.05]{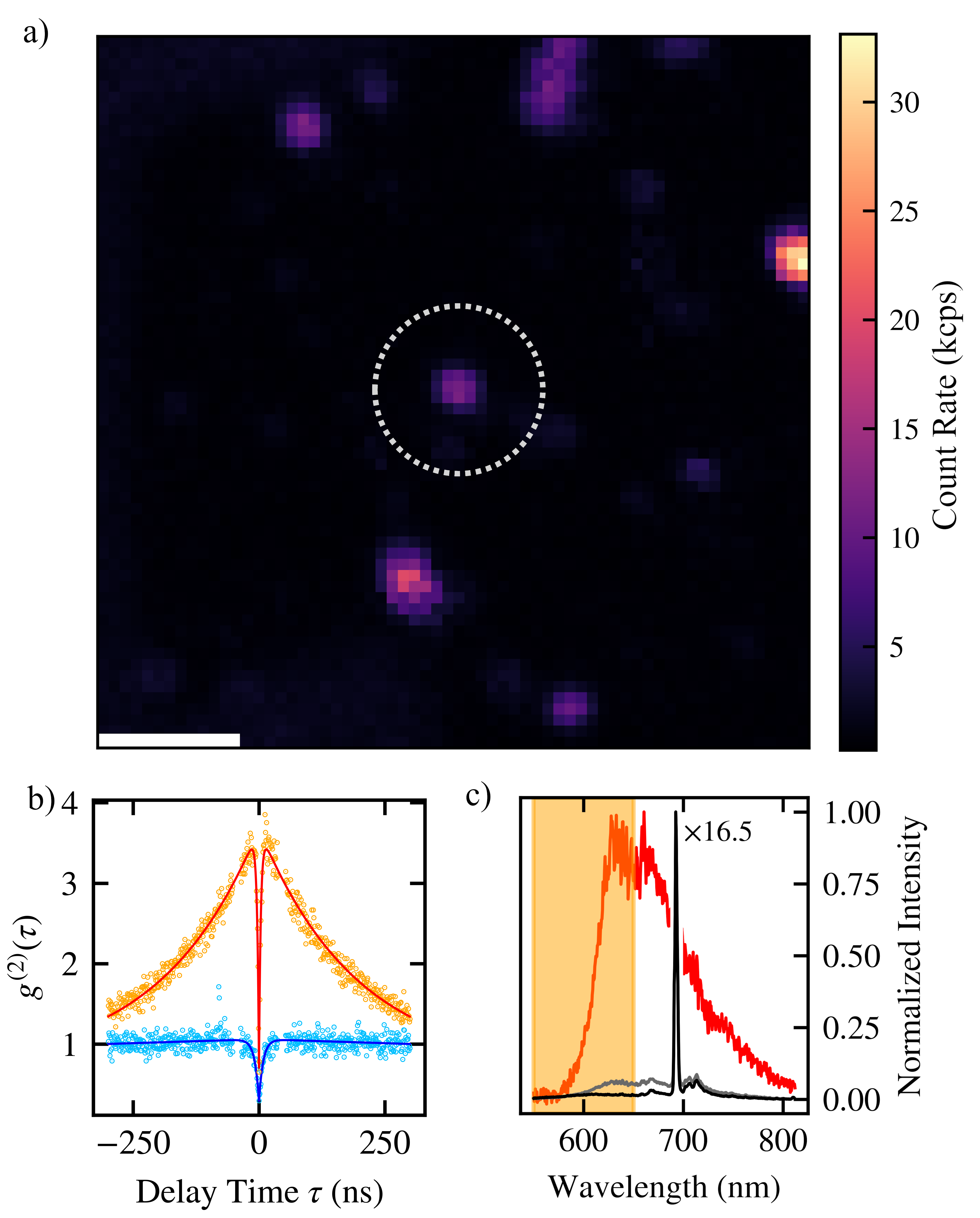}
    \caption{Identifying color centers in aluminum nitride. a) Confocal scan map of an area of the sample, highlighting a representative single color center in the white dashed circle. Scale bar is \SI{2}{\micro \meter}. b) Autocorrelation measurement performed at pump powers of \SI{44.3}{\micro \watt} and \SI{1.40}{\milli \watt}. c) Room temperature spectrum of the color center under \SI{532}{\nano \meter} excitation. The orange region denotes the observation window for photo-physical measurements. The uncorrected spectrum of the CC is given in gray with the spectrum used for the background in black.}
    \label{fig:1}
\end{figure}

The sample measured in this work is a single crystal, \SI{1}{\micro \meter} thick epi-layer of AlN grown via metal-organic chemical vapour deposition on a [0001] plane sapphire substrate. In confocal scan maps, CCs appeared as diffraction limited spots, Figure \ref{fig:1}(a). A confocal microscope was used to excite and collect light from the CCs. The sample was excited with a wavelength of $\lambda_{exc} = $ \SI{532}{\nano\meter} using a DPSS laser. The excitation polarization was purified with an additional linear polarizer. A 0.9 numerical aperture objective focused the excitation laser onto the sample and collected the resultant fluorescence. The fluorescence between \SI{550} and \SI{650}{\nano \meter} was coupled into a SMF28 fibre and subsequently measured with an SPCM-AQRH silicon Avalanche Photo Diode (APD) from Excelitas. The beam was translated laterally on the sample's surface by a mirror galvanometer and 4f imaging system, whilst the focal depth was controlled by a piezoelectric actuator.

The density of CCs is sufficiently low that they can be individually addressed, as shown in Fig. \ref{fig:1}(a). The CC at the middle of the scan has a typical emission spectrum consistent with a zero-phonon line around \SI{620}{\nano \meter} at room temperature (not visible in this example) and a broad phonon side-band extending to 800 nm, shown in Fig. \ref{fig:1}(b). The absence of an obvious zero-phonon line is common amongst other CCs in AlN, indicative of a low Debye-Waller factor \cite{Bishop2020Room-TemperatureNitride, Xue2020Single-PhotonFilms}. We observe the usual indicators of quantised point-like emission, namely resolution limited spot size, saturation of intensity at increasing laser power, and few nanosecond radiative decays in all investigated CCs, as previously reported \cite{Bishop2020Room-TemperatureNitride} (data not shown). Firm proof of quantised electronic states comes in the form of photon statistics displaying anti-bunching under continuous wave excitation with $g^{(2)}(\tau) < 0.5$ at low powers. At low excitation power $g^{(2)}(0) = 0.29 \pm 0.04$. The data is fit with the equation:

\begin{equation}
    g^{(2)}(\tau) = 1-a_{1}e^{-|\tau|/\tau_1} + a_{2}e^{-|\tau|/\tau_2}
\end{equation}
with $\tau_1$ and $\tau_2$ representing the anti-bunching and bunching lifetimes respectively. At \SI{44.3}{\micro \watt} excitation $\tau_1 = 8.3 \pm 0.8$ \si{ns} and $\tau_2 = 3.7 \pm 1.1$ \si{\micro \second}. At \SI{1.40}{\milli \watt} excitation $\tau_1 = 3.5 \pm 0.2$ \si{ns} and $\tau_2 = 198 \pm 7$ \si{ns}. Under stronger optical excitation, increasingly prominent bunching is observed in Fig. \ref{fig:1}(b), suggestive of one or more long-lived shelving states that block photon emission \cite{Kurtsiefer2000StablePhotons}. 

\begin{figure}
    \centering
    \includegraphics{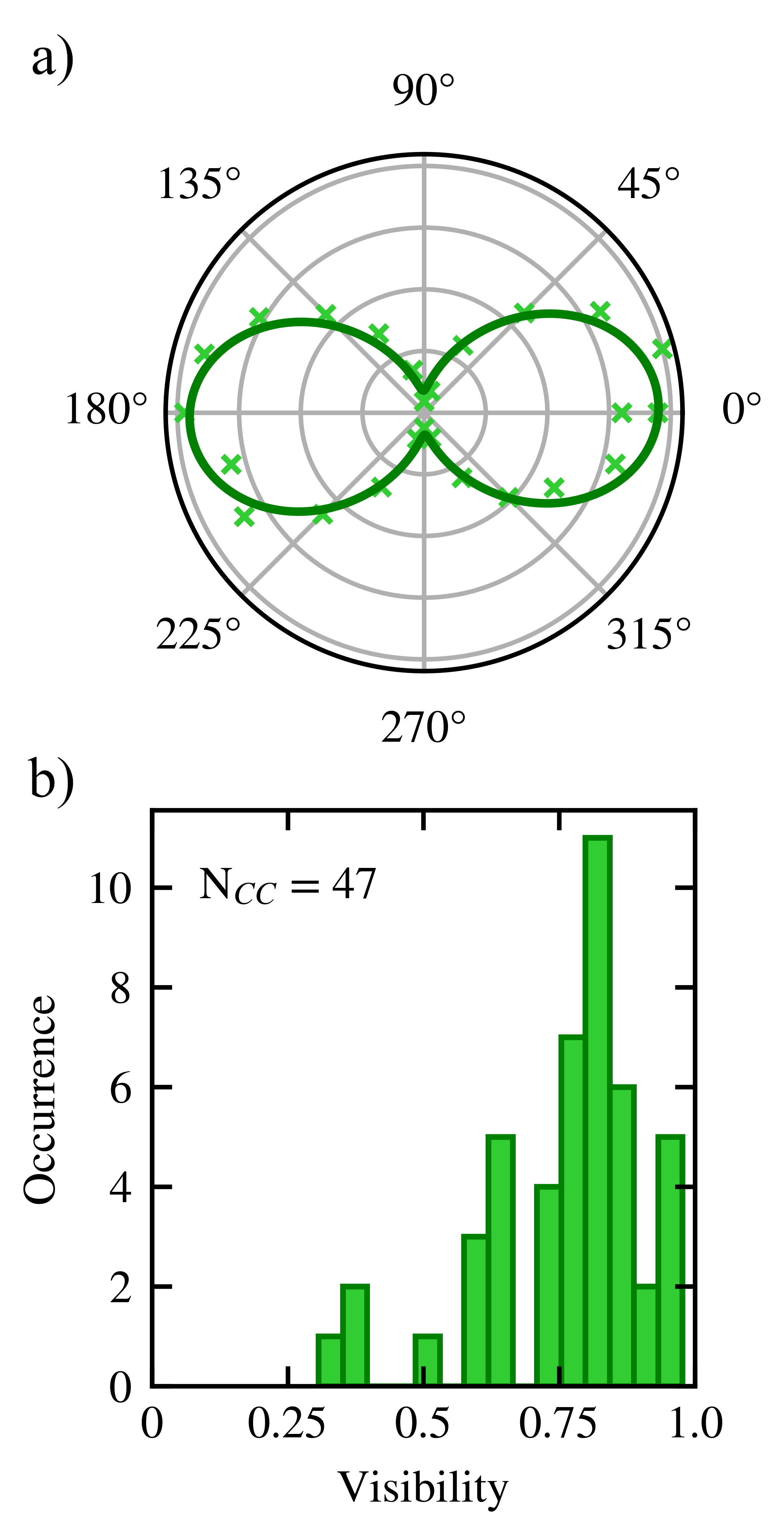}
    \caption{a) Polarization resolved photon counting measurement of a CC performed in absorption. b) Histogram of CC visibilities performed in absorption.}
    \label{fig:2}
\end{figure}

Previous reports of CCs in c-plane AlN have all reported linear absorption dipoles. Motivated to determine whether these absorption dipoles are aligned to the crystal directions, we have developed an automated procedure to measure the polarizations of a large number of CCs. Conventionally, the in-plane orientation angle when maximum absorption occurs, $\theta_{\mathrm{abs}}$, for a given CC is found by continuous rotation of the laser polarization $\theta$ whilst monitoring the emission intensity.  This results in a detected intensity variation, $I(\theta) = a + b\cos^2(\theta - \theta_{\mathrm{abs}}) $ for a given CC located in some initial scan area. $\eta$ is the visibility of the variation in $I(\theta)$. High visibility is consistent with a single linear absorption dipole. Fig \ref{fig:2}(a) illustrates one CC with $\eta = 0.92$. Measurements of 47 CCs are presented in Fig \ref{fig:2}(b) showing the majority of CCs surveyed have a high degree of polarisation, indicating absorption to a single dipole. However, applying this technique to an ensemble of CCs identified in a scan area naturally pre-selects CCs aligned to the laser polarization of the initial scan. A different approach is required to uniformly sample the ensemble. 

\begin{figure*}[ht]
    \centering
    \includegraphics[scale = 1.15]{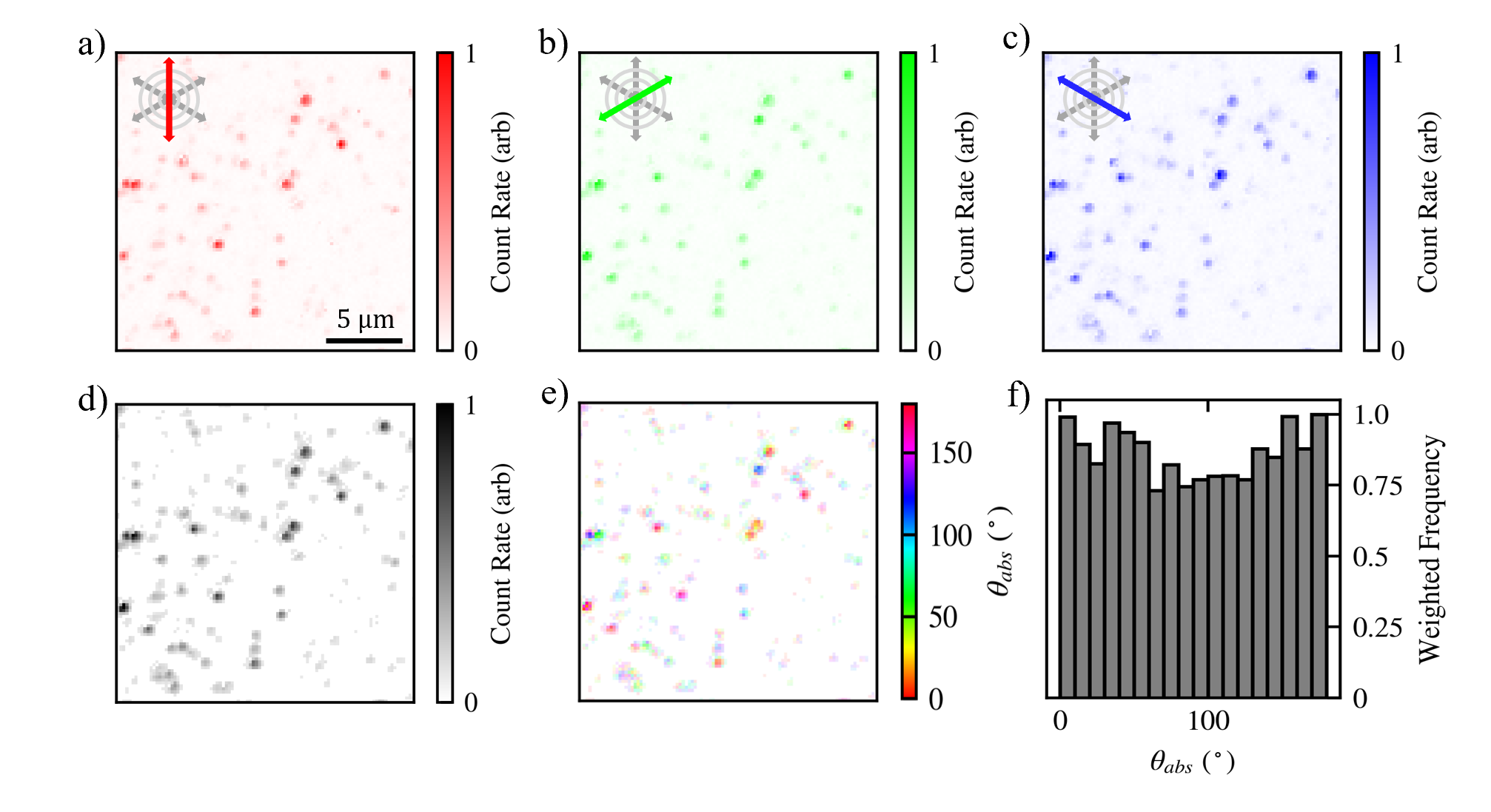}
    \caption{a-c) Room temperature confocal scan maps of the sample at excitation polarization angles of \SI{0}{\degree}, \SI{60}{\degree} and \SI{120}{\degree} using an excitation wavelength of \SI{532}{\nano \meter}. Scale bar is \SI{5}{\micro \meter}. d) Quadrature sum intensity map composed of each pixel in Figures 3a-3c. e) Angle map of the area, weighted by the intensity of the pixels. f) Histogram of pixel orientations as as a function of angle, relative to the crystal. Noise arising from the low intensity areas between pixels is removed by applying an intensity threshold.}
    \label{fig:3}
\end{figure*}

Fig. \ref{fig:3}(a-c) illustrates three scan maps measured over the same area at laser excitation angles $\theta$ = \SI{0}, \SI{60} and \SI{120}{\degree}, without polarization filtering in the collection path. It is evident that some CCs are preferentially excited at one laser polarization, and entirely suppressed in the other scans. To obtain the full picture of the CC locations, the three scan maps in (a)-(c) are combined with the positive-root of their quadrature sum, shown in (d). The CC density is one CC per \SI{11}{\micro \meter \squared}.

It is also possible to build a polarization map for every pixel using the three area maps in Fig. \ref{fig:3}(a-c). This is achieved by defining a vector aligned to the direction of the laser polarization for each angle, $\Vec{I_{0^\circ}}$, $\Vec{I_{60^\circ}}$ and $\Vec{I_{120^\circ}}$, with a magnitude equal to the intensity of each pixel in each map. It can be shown that the vector sum of these maps is aligned to the dipole absorption angle at each pixel. 
\begin{equation}
    \Vec{D_{\mathrm{abs}}} = \sum \left(\Vec{I_{0^\circ}} + \Vec{I_{60^\circ}} + \Vec{I_{120^\circ}} \right)
    \label{eq:vectorSum}
\end{equation}

$\Vec{D_{\mathrm{abs}}}$ is parameterised by its magnitude $r = |\Vec{D_{\mathrm{abs}}}|$ and its angle $\theta_{\mathrm{abs}}$. 

We have verified this method is accurate in measuring the polarization in regions of high intensity, such as where CCs are located, by comparison to subsequent laser polarization scans on individual CCs. However, in the area between CCs the low intensity background contains uncorrelated noise in the $\Vec{I_{0^\circ}}, \Vec{I_{60^\circ}}$ and $\Vec{I_{120^\circ}}$ maps resulting from counting noise in each map. We therefore weight each point in the angle map, Fig. \ref{fig:3}(e), using the corresponding point in the intensity map, Fig. \ref{fig:3}(d). The weighting function, $w(x,y)$, takes the form $w(x,y) = (I(x,y)-I_{\mathrm{Min}})/(I_{\mathrm{Max}}-I_{\mathrm{Min}})$ 
where $I(x,y)$ is the intensity at each location, $I_{\mathrm{Max}}$ the maximum and $I_{\mathrm{Min}}$ the minimum on the intensity map in (d). The resulting angle map in (e) shows polarization as a color, with intensity at each pixel weighted by the intensity in (d). White indicates a low intensity. This procedure provides an efficient and automated method of assessing every CC within an ensemble. 

In our optical system, the birefringence of a dichroic beamsplitter results in a small reduction in the purity of the laser polarization at some angles. The maximum visibility achievable by the microscope is \SI{99.8}{\%}. When the effect of the birefringence was greatest, the visibility was \SI{94.8}{\%}, which has no impact on the conclusions of our study.

With this information we can look at the statistics of CC absorption angles over the \SI{2025}{\micro \meter \squared} area. A histogram of weighted pixel angle, shown in Fig. \ref{fig:3}(f), indicates that the absorption dipole angles of CCs are uniformly distributed. This in contrast to many other solid state emitters, for instance the neutral exciton in InAs quantum dots usually has two dipoles along crystal axes in the plane of the sample \cite{Bennett2010} and the NV$^-$ in diamond has four possible orientations which can be readily observed in [111] oriented samples \cite{Alegre2007Polarization-selectiveDiamond}. We anticipate that scanning samples featuring CCs with high $\eta$ and a limited number of dipole  orientations would yield distinct peaks at angles corresponding to the crystal axis of the sample.

The lack of any preferred direction in the in-plane component of absorption dipoles in AlN CCs points to a more complex origin than the diamond NV$^-$. In future, sampling the ensemble distribution of emission dipole orientations could yield further insight into the physical origin of these CCs, or could be applied to other material systems. Identification of all CCs in these samples, not just those aligned to the laser in a single scan, could be useful in correlative microscopy techniques where transmission electron microscopy and cathodoluminescence is compared to PL maps to investigate the link between threading dislocations in AlN and CCs.

\section*{Acknowledgments}

We acknowledge financial support provided by EPSRC via Grant No. EP/T017813/1 and the European Union's H2020 Marie Curie ITN project LasIonDef (GA No. 956387). Device processing was carried out in the cleanroom of the ERDF-funded Institute for Compound Semiconductors (ICS) at Cardiff University.

\section*{Data Availability}

Data supporting the findings of this study are available in the Cardiff University Research Portal at http://doi.org/10.17035/d.2023.0248563719.

\section*{References}
\bibstyle{naturemag}
\bibliography{references}

\end{document}